\documentclass[showpacs,10pt,twocolumn,prb]{revtex4-1}
\usepackage{amsmath}
\usepackage{amssymb}
\usepackage{graphics}
\usepackage{epsfig}
\usepackage{color}

\begin{document}


\title{Superconductivity in kagome metal YRu$_{3}$Si$_{2}$ with strong electron correlations}
\author{Chunsheng Gong, Shangjie Tian, Zhijun Tu, Qiangwei Yin, Yang Fu, Ruitao Luo, and Hechang Lei}
\email{hlei@ruc.edu.cn}
\affiliation{Department of Physics and Beijing Key Laboratory of Opto-electronic Functional Materials $\&$ Micro-nano Devices, Renmin University of China, Beijing 100872, China}

\date{\today}

\begin{abstract}

We report the detailed physical properties of YRu$_{3}$Si$_{2}$ with the Ru kagome lattice at normal and superconducting states.
The results of resistivity and magnetization show that YRu$_{3}$Si$_{2}$ is a type-II bulk superconductor with $T_{c}\sim$ 3.0 K.
The specific heat measurement further suggests that this superconductivity could originate from the weak or moderate electron-phonon coupling.
On the other hand, both large Kadawaki-Woods ratio and Wilson ratio indicate that there is a strong electron correlation effect in this system, which may have a connection with the featured flat band of kagome lattice.

\end{abstract}

\begin{center}
\end{center}
\maketitle


\section{Introduction}

Due to the unique two-dimensional (2D) structure formed by corner-sharing triangles, the kagome lattice provides an exciting platform for studying magnetic frustration, electronic correlation and topological electronic state.
Besides the long-sought spin liquid state in insulating magnetic kagome materials with strong geometrical frustration \cite{Balents,Broholm,Shores,Han}, the topological electronic structure in metallic kagome systems (kagome metals), such as Dirac point, flat band and von Hove point rooting in the special symmetry and arrangement of 2D kagome lattice, has also attracted much attention recently \cite{YeL,KangM,LiuZ,KangM2}.
More importantly, the interplay between magnetism and band topology has resulted in the emergence of various exotic correlated topological phenomena and matters, such as giant anomalous Hall effect, massive Dirac fermions with large magnetic field tunability, magnetic Weyl semimetal state, and Chern gapped Dirac fermions with chiral edge state \cite{Nakatsuji,Nayak,YeL,WangQ1,LiuE,WangQ2,Yin1,Yin2}.

In contrast, the experimental studies on other electronic correlation effects intertwining with band topology in kagome metals are scarce due to the lack of material systems.
The discovery of superconductivity and CDW state in $A$V$_{3}$Sb$_{5}$ ($A$ = K, Rb, Cs) in very recent provide a novel platform to investigate such relationship \cite{Ortiz,Ortiz2,Ortiz3,YinQW}, which has been intensively studied in theory previously \cite{Ko,WangWS,Kiesel,Mazin}.
Moreover, the observed chiral charge order at high temperature and pair density wave state below superconducting transition temperature $T_{c}$ further imply the intricate relationship between these correlated ordering states and non-trivial band topology \cite{JiangYX,ChenH}.	

In order to deepen the understanding of such correlated topological phenomena, it is necessary to explore other kagome metals harbouring superconductivity (kagome superconductors) but only a handful of kagome superconductors have been reported until now. One of examples is the Laves phase [cubic MgCu$_{2}$ (C15) type] superconductors with widely varying $T_{c}$ ranging from 0.07 K to above 10 K \cite{Rapp,SunS,Gong}.
However, since the strong interaction between kagome layer and other layers along the (111) direction of cubic Laves phase, the featured electronic structure of 2D kagome lattice is strongly disturbed and the topological features of these materials become hard to discern.
Another family of kagome superconductors is $RT$$_{3}$B$_{2}$ ($R$ = Y, La, Lu, Th and $T$ = Ru, Os, Rh, Ir) and $R$Ru$_{3}$Si$_{2}$ with hexagonal layered CeCo$_{3}$B$_{2}$ structure in which transition metals form kagome layers \cite{Ku}.
Among these materials, LaRu$_{3}$Si$_{2}$ with $T_{c}$ as high as 7.8 K attracted lots of attention because it shows a typical band structure of kagome lattice and strong electron correlations which could be closely related to the kagome flat band \cite{Barz,Escorne,Li,BXLi,III}.
When compared with LaRu$_{3}$Si$_{2}$, except the structure and $T_{c}$ value ($\sim$ 3 K - 3.5 K) \cite{Barz,Vandenberg}, the physical properties of YRu$_{3}$Si$_{2}$, especially superconducting properties, are still lacking.
In this work, we carry out a detailed study on the physical properties of YRu$_{3}$Si$_{2}$ at normal and superconducting states.
Experimental results indicate that YRu$_{3}$Si$_{2}$ is a weakly or moderately coupled BCS superconductor with strong electron correlations.

\section{Experimental}

YRu$_{3}$Si$_{2}$ polycrystal was synthesised using an arc-melting method. 
First, Y metal filing (99.9 $\%$), Ru powder (99.9 $\%$), and Si powder (99.9 $\%$) with stoichiometric ratios were mixed, ground thoroughly and then pressed into a pellet using a hydraulic press in a glover box filled with argon atmosphere. In order to avoid the formation of YRu$_{2}$Si$_{2}$, an amount of extra Ru was added, similar to the preparation of LaRu$_{3}$Si$_{2}$ \cite{Li}. The pellet was arc-melted under argon atmosphere and remelted several times from both sides of the pellet to improve the homogeneity.
Crystal structure and phase purity were examined by powder X-ray diffraction pattern (PXRD) with Cu K$_{\alpha }$ radiation ($\lambda $ = $0.15418$ nm) using a Bruker D8 X-ray diffractometer. The lattice parameters are extracted by fitting the PXRD pattern using the TOPAS4 software.$\cite{topas}$
Electrical transport and specific heat measurements were performed in a Quantum Design PPMS-14T. Magnetization measurements were carried out in a Quantum Design MPMS3.

\section{Results and discussion}

\begin{figure}[tbp]

\centerline{\includegraphics[scale=0.24]{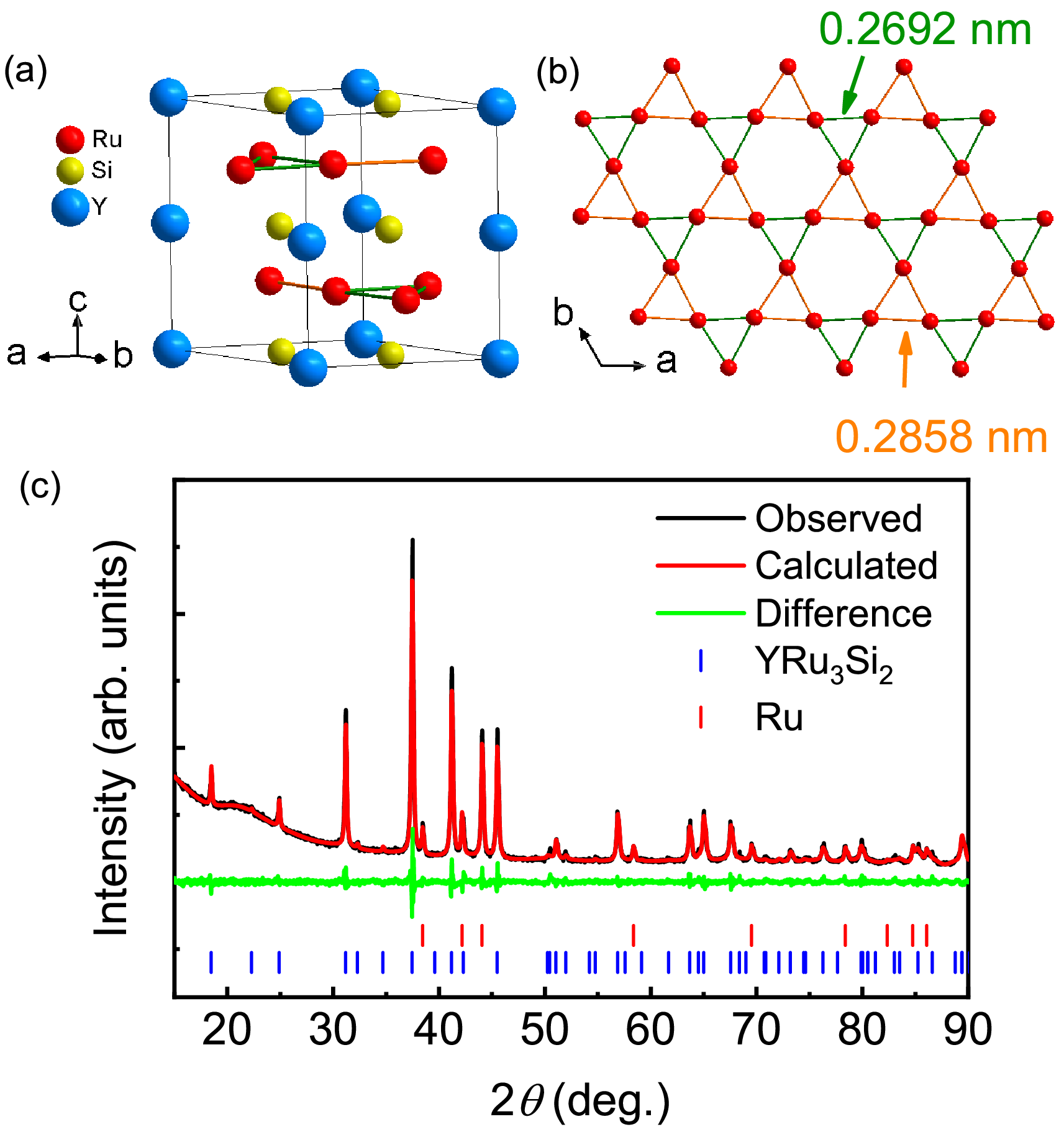}} \vspace*{-0.5cm}
\caption{(a) Crystal structure of YRu$_{3}$Si$_{2}$. The blue, red and yellow balls represent Y, Ru and Si atoms, respectively. (b) Top view of the 2D distorted kagome lattice of Ru atoms. Two different bond distances are labelled with green and orange lines. (c) Powder XRD pattern and Rietveld refinement of YRu$_{3}$Si$_{2}$ polycrystal.}
\end{figure}

Fig. 1(a) shows the crystal structure of YRu$_{3}$Si$_{2}$, isostructural to LaRu$_{3}$Si$_{2}$ \cite{Vandenberg}.
It is formed by stacking Y-Si and Ru layers alternately along the $c$ axis.
A key feature of this compound is the 2D distorted kagome layer formed by Ru atoms parallel to the $ab$ plane.
In this layer, there are two different Ru-Ru bond distances $d_{\rm Ru-Ru}$ (= 0.2692 nm and 0.2858 nm). Meanwhile, the rotation of Ru triangles leads to the bond angle for collinear Ru atoms $\theta$ (= 174.046$^{\circ}$) deviating away from $\theta=$ 180$^{\circ}$ in perfect kagome lattice (Fig. 1(b)).
On the other hand, for Y-Si layer, the Y atoms form a triangle lattice when the Si atoms construct a honeycomb structure.
Fig. 1(c) shows the PXRD pattern of YRu$_{3}$Si$_{2}$ and all peaks can be indexed well by the $P6_{3}/m$ space group (No. 176). The obtained lattice parameters by using Rietveld refinement are $a=$ 0.5542(1) nm and $c=$ 0.7150(7) nm, consistent with the reported values in literature $\cite{Vandenberg}$.
In addition to the diffraction peaks of YRu$_{3}$Si$_{2}$, there are some extra peaks which originate from the excess of Ru in the raw material. A fitting result shows that the weight ratio of YRu$_{3}$Si$_{2}$ to Ru is about 84.1 : 15.9.

\begin{figure}[tbp]
\centerline{\includegraphics[scale=0.35]{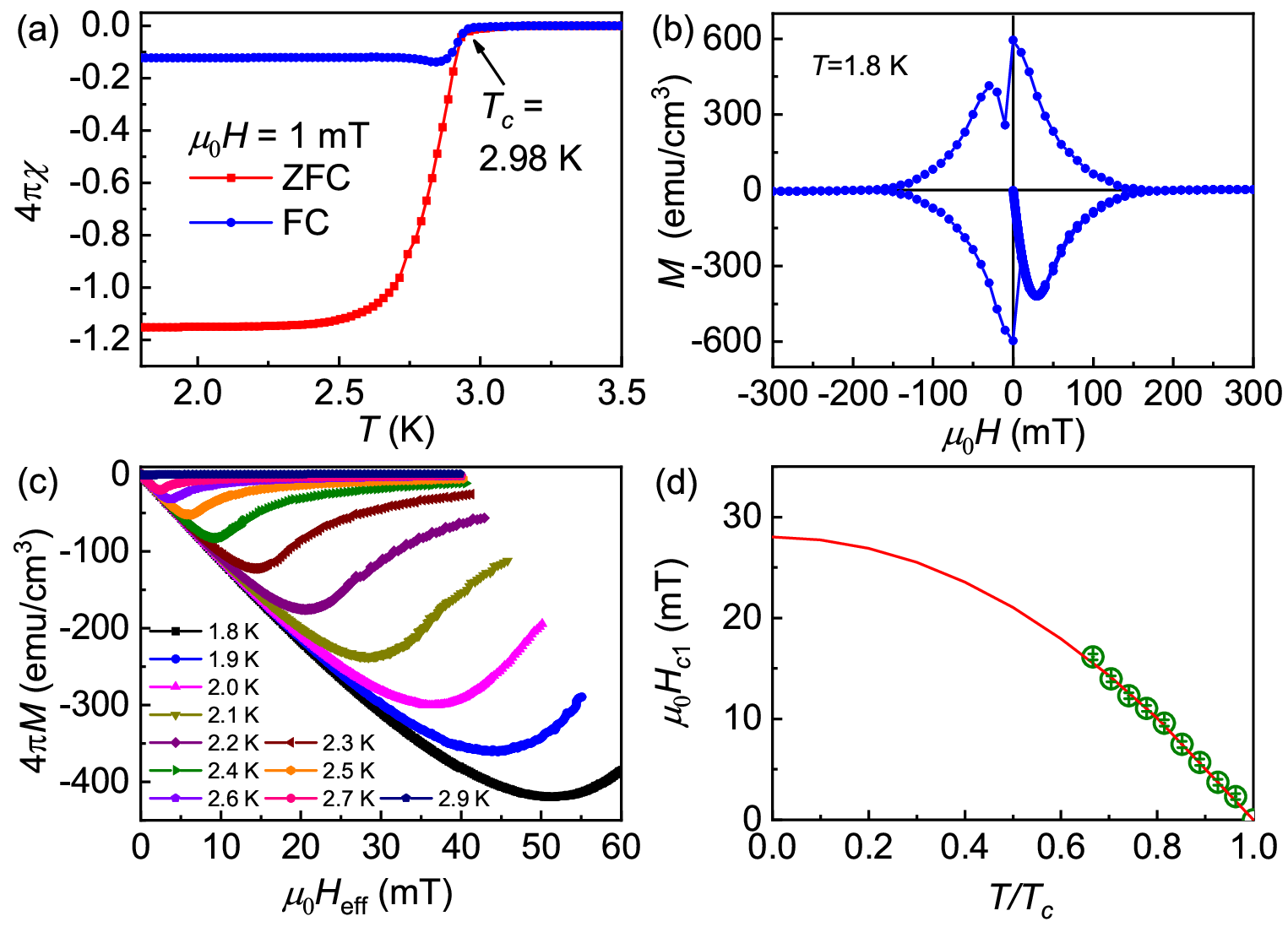}} \vspace*{-0.5cm}
\caption{(a) Temperature dependence of magnetic susceptibility $4\pi\chi(T)$ at 1 mT with ZFC and FC models. (b) Magnetization hysteresis loop for YRu$_{3}$Si$_{2}$ at 1.8 K. (c) Low-field dependence of magnetization $4\pi M(\mu_{0}H)$ at various temperatures below $T_{c}$. (d) $\mu_{0}H_{c1}$ as a function of $T/T_{c}$. The red line represents the fit using G-L equation.}
\end{figure}

The temperature dependent magnetic susceptibility $4\pi\chi(T)$ with the zero field cooling (ZFC) and field cooing (FC) modes is shown in Fig. 2(a). A clear diamagnetic transition in $4\pi \chi(T)$ curve can be observed and confirms the occurrence of superconductivity in YRu$_{3}$Si$_{2}$. The $T_{c}$ determined from the ZFC $4\pi \chi(T)$ curve at $\mu_{0}H=$ 1 mT is about 2.98 K, close to the value reported previously \cite{Barz}.
At $T=$ 1.8 K, the estimated superconducting volume fraction is about 110 \%, indicating a bulk superconductivity in YRu$_{3}$Si$_{2}$.
Compared with the ZFC $4\pi \chi(T)$ curve, a relatively weak diamagnetic signal of FC $4\pi \chi(T)$ curve due to flux pinning effect suggests that YRu$_{3}$Si$_{2}$ is a type-II superconductor. It is further confirmed by the magnetization hysteresis loop measured at 1.8 K (Fig. 2(b)).
Moreover, the sudden jumps of $M(\mu_{0}H)$ at low-field region can be ascribed to the entry of Meissner state.
The initial magnetization as a function of magnetic field in the temperature range between 1.8 K and 2.9 K is shown in Fig. 2(c).
All curves clearly fall on the same line and deviate from linearity at different temperatures.
In order to estimate the lower critical field $\mu_{0}H_{c1}$ correctly, the effective field $\mu_{0}H_{\rm eff}$ is calculated by using the formula $\mu_{0}H_{\rm eff} =\mu_{0}H$ - $N_{d}M$, where $N_{d}$ is the demagnetization factor \cite{Aharoni}. Using the geometry size of rectangular sample, the calculated value of $N_{d}$ is about 0.522 \cite{Aharoni}.
The $\mu_{0}H_{c1}$ at each temperature is determined from the field where the $4\pi M(\mu_{0}H)$ curve deviates from linearity ("Meissner line").
The obtained $\mu_{0}H_{c1}$ as a function of reduced temperature $T/T_{c}$ is shown in Fig. 2(d).
The temperature dependence of $\mu_{0}H_{c1}$ can be well fitted by the Ginzburg-Landau (G - L) equation $\mu_{0}H_{c1}(T) = \mu_{0}H_{c1}(0) [1-({T}/{T_{c}})^{2}]$, where $\mu_{0}H_{c1}(0)$ is the lower critical field at $T=$ 0 K (red line). The fitted $\mu_{0}H_{c1}(0)$ for YRu$_{3}$Si$_{2}$ is 28.0(3) mT.

\begin{figure}[tbp]
\centerline{\includegraphics[scale=0.55]{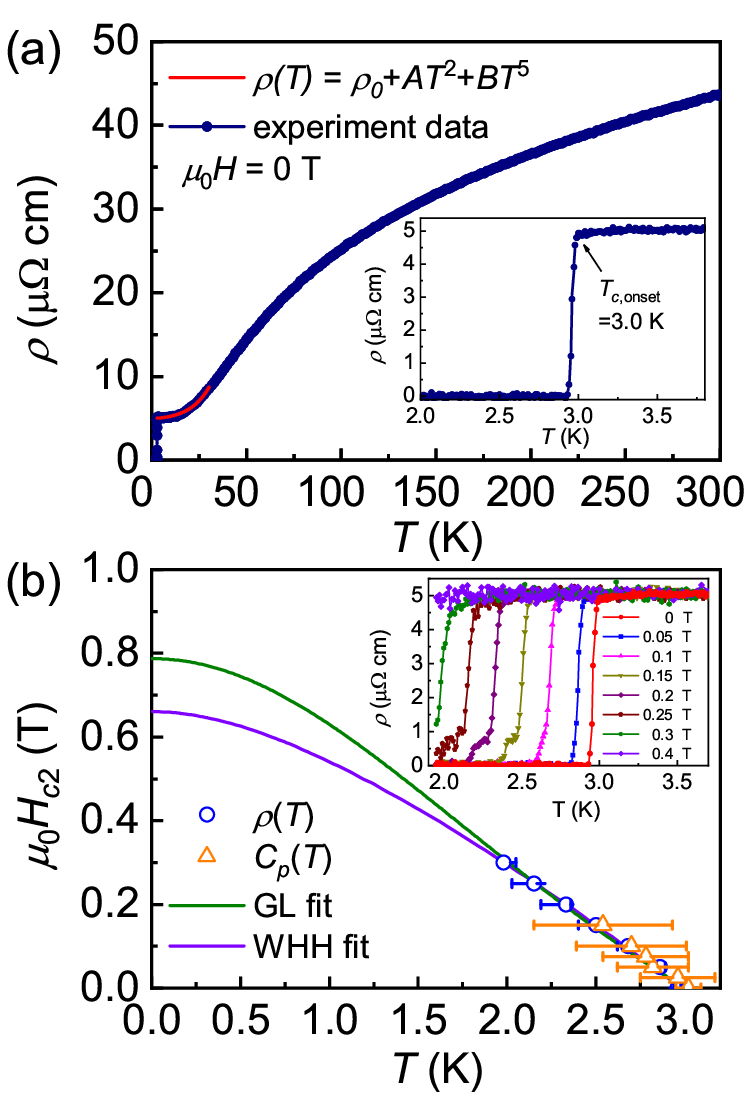}} \vspace*{-0.3cm}
\caption{(a) Temperature dependence of electrical resistivity $\rho(T)$ for YRu$_{3}$Si$_{2}$ polycrystal at zero field. The red solid line represents the fit using the formula $\rho(T)=\rho_{0}+AT^2+BT^{5}$. Inset: enlarged view of $\rho(T)$ curve near superconducting transition. (b) Temperature dependence of $\mu_{0}H_{c2}(T)$. The blue circles and orange triangles represent the $\mu_{0}H_{c2}(T)$ determined from resistivity and specific heat measurements, respectively. The error bars for the former are determined from the 10\% and 90\% of normal state resistivity just above superconducting transition. The error bars for the later are determined from the starting points and peak positions of jumps on the $C_{p}(T)$ curves. The green and purple lines represent the fits using the G - L and WHH formulas. Inset: $\rho(T)$ as a function of $T$ at various magnetic fields.}
\end{figure}

Fig. 3(a) shows the temperature dependence of electrical resistivity $\rho(T)$ for YRu$_{3}$Si$_{2}$ polycrystal from 300 K to 2 K at zero field. The $\rho(T)$ decreases monotonically with decreasing temperature, indicating the metallic behavior of YRu$_{3}$Si$_{2}$.
The value of residual resistivity ratio (RRR) $\rho$(300 K)/$\rho$(4 K) is about 8.6. Such relatively small value of RRR can be explained by the grain boundary effect in YRu$_{3}$Si$_{2}$ polycrystal.
At high temperature the $\rho(T)$ curve exhibits a saturation tendency. This behavior could be described by the Ioffe-Regel limit\cite{Ioffe}, e.g. the electron mean free path is close to the interatomic distance \cite{Zverev}.
At low temperature region ($T<$ 30 K), the $\rho(T)$ curve can be well fitted using the formula $\rho(T)=\rho_{0}+AT^2+BT^{5}$, where $\rho_{0}$ is the residual resistivity, $AT^{2}$ and $BT^{5}$ term originates from the electron-electron and electron-phonon scattering, respectively  (red line). The fit gives $\rho_{0}=4.99(5)$ $\mu\Omega$ cm, $A=$ 2.38(7)$\times10^{-3}$ $\mu\Omega$ cm K$^{-2}$ and $B=$ 5.9(4)$\times10^{-8}$ $\mu\Omega$ cm K$^{-5}$.
In addition, because the value of $AT^{2}$ is always larger than $BT^{5}$ at $T<$ 30 K, the electron-electron scattering process dominates low-temperature resistivity.
With lowering temperature further, a sharp superconducting transition can be observed at $T_{c,\rm onset}$ = 3.0 K with narrow transition width $\Delta T_{c}$ = 0.082 K (inset of Fig. 3(a)). It is consistent with $T_{c}$ obtained from the $4\pi\chi(T)$ curve. The small value of $\Delta T_{c}$ indicates the high quality of YRu$_{3}$Si$_{2}$ polycrystal.


In order to investigate the upper critical field $\mu_{0}H_{c2}$ in YRu$_{3}$Si$_{2}$, the temperature dependence of $\rho(T)$ at various fields up to 0.4 T are measured (inset of Fig. 3(b)). The $T_{c}$ shifts to lower temperature and the $\Delta T_{c}$ broadens slightly with increasing field.
At $\mu_{0}H=$ 0.4 T, the superconducting transition can not be observed above 2 K.
When defining the $T_{c}$ as 50\% drop from normal state resistivity just above superconducting transition, the temperature dependence of $\mu_{0}H_{c2}(T)$ for YRu$_{3}$Si$_{2}$ is plotted in Fig. 3(b).
Using the G - L formula $\mu_{0}H_{c2}(T) = \mu_{0}H_{c2}(0)(T_{c,0}^{2}-T^{2})/(T_{c,0}^{2}+T^{2})$, where the $\mu_{0}H_{c2}(0)$ is the zero-temperature upper critical field and $T_{c,0}$ is superconducting transition temperature at zero field. The fit (green line) gives $\mu_{0}H_{c2}(0)=$ 0.78(2) T.
On the other hand, when using the one-band Werthamer-Helfand-Hohenberg (WHH) formula (purple line) \cite{Werthamer}, the fitted $\mu_{0}H_{c2}(0)$ is 0.655(2) T, close to the value obtained from the G-L formula.
It is lower than that of LaRu$_{3}$Si$_{2}$ ($\mu_{0}H_{c2}(0) \approx$ 4 T)\cite{Li}.
In addition, the $\mu_{0}H_{c2}(0)$ is also much smaller than the Pauli paramagnetically limited field $\mu_{0}H^{p}_{c2}$ (= 1.84 $T_{c}$ = 5.52 T) \cite{Maki}, indicating that the orbital pairing-broken mechanism is dominant in YRu$_{3}$Si$_{2}$.

Using the fitted values of $\mu_{0}H_{c1}(0)$ (= 28.0(3) mT) and $\mu_{0}H_{c2}(0)$ (= 0.655(2) T) , the zero-temperature superconducting characteristic parameters $\xi_{\rm GL}$ (coherence length) and $\lambda_{\rm GL}$ (penetration depth) can be estimated according to the following two equations $\xi_{\rm GL} = (\frac{\phi_{0}}{2\pi\mu_{0}H_{c2}(0)})^{\frac{1}{2}}$ and $\mu_{0}H_{c1} =\frac{\phi_{0}}{4\pi\lambda^{2}_{\rm GL}}\ln\frac{\lambda_{\rm GL}}{\xi_{\rm GL}}$, where $\phi_{0}$ is the magnetic flux quantum ($h/2e$ = 2.07 $\times10^{-15}$ T m$^{2}$).
The calculated value of $\xi_{\rm GL}$ and $\lambda_{\rm GL}$ is 22.43 (2) nm and 24.80(3) nm, respectively.
Correspondingly, the derived GL parameter $\kappa_{\rm GL} (= \lambda_{\rm GL}/\xi_{\rm GL}$) is 1.11(4), which is larger than 1/$\sqrt{2}$. It confirms that YRu$_{3}$Si$_{2}$ is a type-II superconductor. The zero-temperature thermodynamic critical field $\mu_{0}H_{c}(0)$ is obtained to be 0.42(9) T using the equation $\mu_{0}H_{c}(0)=[\mu_{0}H_{c1}(0)\mu_{0}H_{c2}(0)/\ln \kappa_{\rm GL}]^{1/2}$.


\begin{figure}[tbp]
\centerline{\includegraphics[scale=0.35]{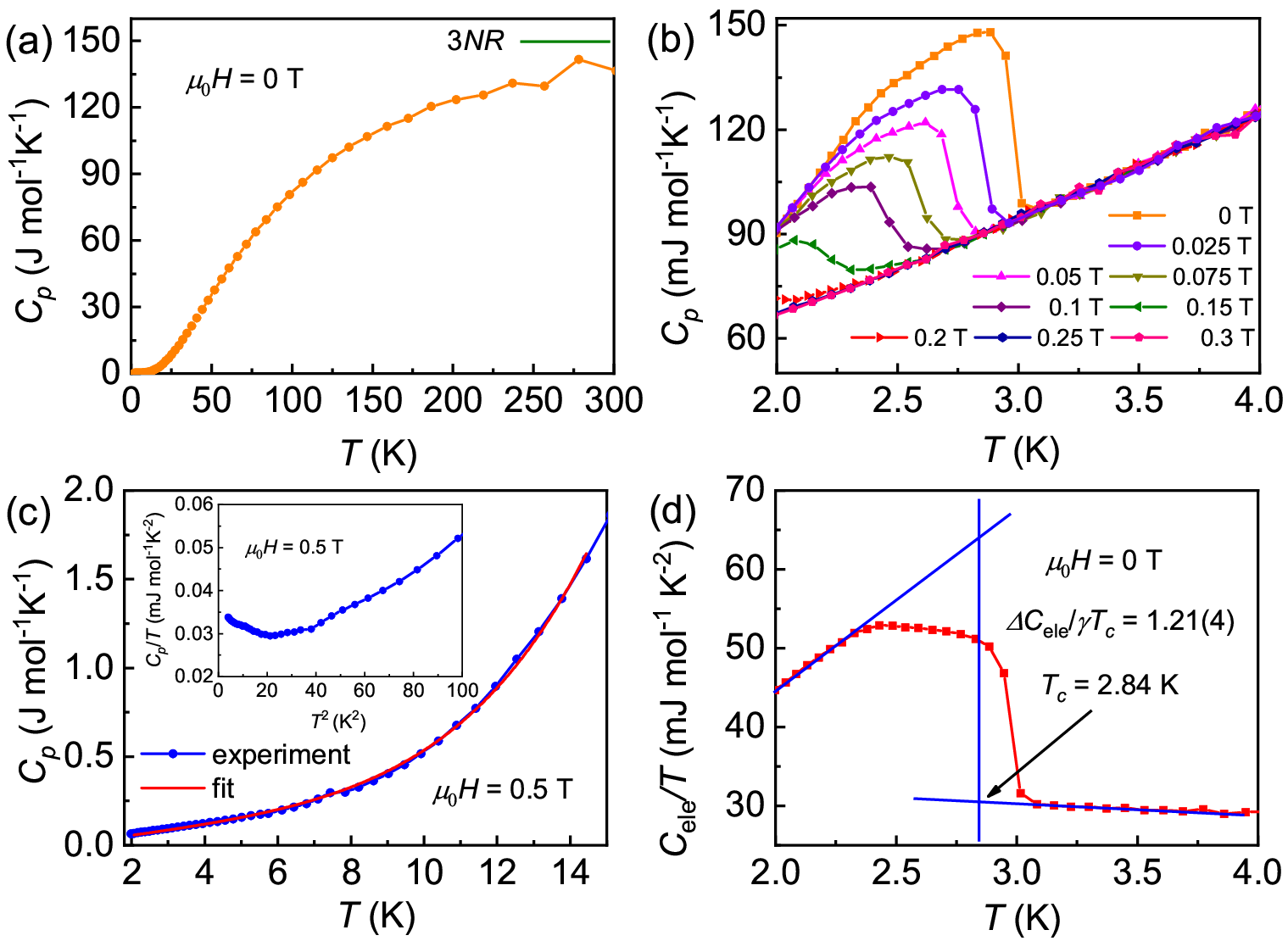}} \vspace*{-0.3cm}
\caption{(\textrm{a}) Temperature dependence of zero-field $C_{p}(T)$ from 300 K to 2 K. (b) $C_{p}$ vs $T$ at various magnetic fields. (c) Temperature dependence of $C_{p}(T)$ at $\mu_{0}H$ = 0.5 T. The red solid line represents the fit using the formula $C_{p}=\gamma T+\beta T^3+\eta T^5$. Inset: ${C_{p}}/{T}$ vs $T^{2}$ at $\mu_{0}H$ = 0.5 T. (d) ${C_{\rm ele}}/{T}$ as a function of $T$ at zero field.}
\end{figure}

Fig. 4(a) shows the specific heat $C_{p}(T)$ of YRu$_{3}$Si$_{2}$ measured from 300 K to 2 K at zero field.
At high temperature, consistent with the Dulong-Petit law, the $C_{p}(T)$ is close to the value of 3$NR$ ($\sim$ 150 J mol$^{-1}$K$^{-1}$), where $N$ (= 6) is the atomic numbers per formula and $R$ (= 8.314 J mol$^{-1}$K$^{-1}$) is the ideal gas constant, respectively.
At low temperature, a specific heat jump can be observed clearly due to the superconducting transition. It also confirms the bulk nature of superconductivity in YRu$_{3}$Si$_{2}$ (Fig. 4(b)). With increasing field, this jump shifts to lower temperature accompanying with the decrease of height.
On the other hand, at $\mu_{0}H$ = 0.5 T where the superconducting transition is suppressed below 2 K, the relationship of $C_{p}/T$ and $T^2$ can not described by a linear behavior (inset of Fig. 4(c)) and this suggests that there may be an anomalous contribution of high-frequency phonons \cite{Li}.
Thus, the low-temperature $C_{p}(T)$ at $\mu_{0}H$ = 0.5 T is fitted using the formula $C_{p}=\gamma T+\beta T^3+\eta T^5$, where the first item is the specific heat of electrons at normal state with the Sommerfeld coefficient $\gamma$, and the last two items represent the lattice specific heat when considering the contribution of high-frequency phonons \cite{Yang}.
The fit gives $\gamma$ = 27.5(8) mJ mol$^{-1}$K$^{-2}$, $\beta$ = 0.12(2) mJ mol$^{-1}$K$^{-4}$ and $\eta$ = 0.00141(7) mJ mol$^{-1}$K$^{-6}$. By using the formula $\Theta_{\rm D} =({12 \pi ^4 NR}/{5 \beta})^{1/3}$, the calculated Debye temperature $\Theta_{\rm D}$ is 460(26) K. It is larger than the value in LaRu$_{3}$Si$_{2}$ ($\Theta_{\rm D}=$ 412 K) derived using the same method \cite{Li}, possibly due to the smaller atomic mass of Y than La.

The specific heat of electrons $C_{\rm ele}$ can be obtained by subtracting the phonon contribution from the total specific heat (Fig. 4(d))
and the phonon contribution is obtained from the fit of $C_{p}(T)$ curve measured at 0.5 T. According to the method of equal-entropy construction (blue solid lines), the thermodynamic $T_{c}$ is determined to be 2.84 K, consistent with the values obtained from the $\rho(T)$ and $\chi(T)$ curves. In addition, the determined $\mu_{0}H_{c2}(T)$ from the $C_{p}(T)$ curves at various fields also exhibits similar trend when compared to the $\mu_{0}H_{c2}(T)$ data derived from the resistivity measurements (orange triangles in Fig. 3(b)).
On the other hand, the specific heat jump ${\Delta C_{\rm ele}}/{\gamma T_{c}}$ at $T_{c}$ is 1.21(4) which is smaller than the value of weak-coupling limit 1.43.
It implies that YRu$_{3}$Si$_{2}$ might be a weak- or moderate-coupling superconductor.
It has to be noted that the existence of Ru impurity could also contribute to the reduced specific heat jump.
Using the obtained $\Theta_{\rm D}$ and $T_{\rm c}$, the electron-phonon coupling constant $\lambda_{\rm e-ph}$ can be calculated according to the McMillan equation\cite{McMillan},

\begin{equation}
\lambda_{\rm e-ph} = \frac{1.04+\mu^* \ln (\frac{\Theta_{\rm D}}{1.45 T_{c}})}{(1-0.62 \mu^*) \ln (\frac{\Theta_{\rm D}}{1.45 T_{c}})-1.04}
\end{equation}

where $\mu^*$ is the Coulomb pseudopotential parameter. Taken $\mu^*=$ 0.13 as for many intermetallic superconductors, the calculated value of $\lambda_{\rm e-ph}$ is 0.50(2), further confirming that YRu$_{3}$Si$_{2}$ is a weak- or moderate-coupling superconductor. All of physical parameters of YRu$_{3}$Si$_{2}$ are listed in Table I.

\begin{table}[tbp]\centering
\caption{Physical parameters of YRu$_3$Si$_{2}$ at superconducting and normal states.}
\begin{tabular}{ccc}
		\hline\hline
		Parameter                              &Units                                          &Value                     \\\hline
		$T_{c}$                                &K                                              &3.0                       \\
		$\mu_{0}H_{c1}(0)$                     &mT                                             &28.0(3)                   \\
		$\mu_{0}H_{c2}(0)$                     &T                                              &0.655(2)                  \\
		$\mu_{0}H_{c}(0)$                      &T                                              &0.42(9)                   \\
		$\xi_{\rm GL}$                         &nm                                             &22.43(2)                  \\
		$\lambda_{\rm GL}$                     &nm                                             &24.80(3)                  \\
		$\kappa_{\rm GL}$                      &-                                              &1.11(4)                   \\
		${\Delta C_{\rm ele}}/{\gamma T_{c}}$  &-                                              &1.21(4)                   \\		
		$\lambda_{\rm e-ph}$                   &-                                              &0.50(2)                   \\	
		$\gamma$                               &mJ mol$^{-1}$K$^{-2}$                          &27.5(8)                   \\
		$\Theta_{\rm D}$                       &K                                              &460(26)                   \\
        $A/\gamma^{2}$                         &$\mu \Omega$ cm mol$^{2}$ K$^{2}$ J$^{-2}$     &3.15(2)                   \\
		$R_{\rm w}$                            &-                                              &1.49(4)                   \\
		\hline\hline
\end{tabular}
\label{Table 1}
\end{table}

%


\begin{figure}[tbp]
\centerline{\includegraphics[scale=0.3]{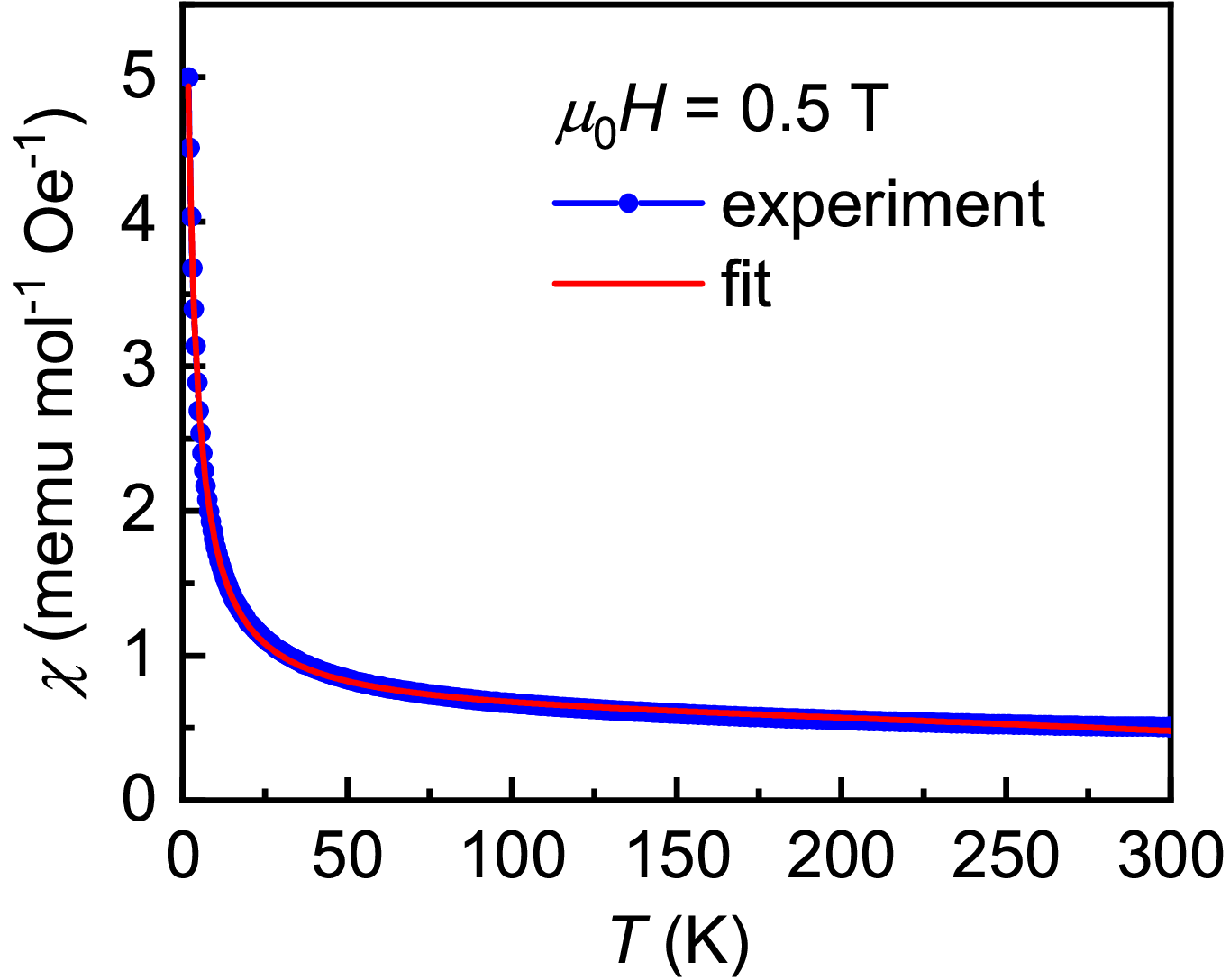}} \vspace*{-0.3cm}
\caption{Temperature dependence of $\chi(T)$ measured at $\mu_{0}H$ = 0.5 T with ZFC mode. The red solid line presents the fit using the formula $\chi(T)$ = $\chi(0)[1-(T/T_{\rm E})^{2}]+C/(T+T_{0})$.}
\end{figure}

In order to get more information about electron correlation effect in YRu$_{3}$Si$_{2}$, temperature dependence of $\chi(T)$ at $\mu_{0}H$ = 0.5 T with ZFC mode is measured (Fig. 5). YRu$_{3}$Si$_{2}$ shows a paramagnetic behavior from 300 K to 2 K and no long-range magnetic translation is observed. In addition, it is found that the $\chi(T)$ curve can be well fitted using the following expression as in LaRu$_{3}$Si$_{2}$ \cite{Li},

\begin{equation}
\chi(T) = \chi(0)[1-(\frac{T}{T_{\rm E}})^{2}]+\frac{C}{T+T_{0}}
\end{equation}

Here, the first term represents the Pauli susceptibility related to the density of state at the Fermi energy level $E_{\rm F}$. The $T_{E}$ represents a parameter proportional to the $E_{\rm F}$. The second term refers to the weak magnetic contribution that may be caused by the local moment. The fit gives $\chi(0)$ = 5.56(2)$\times$10$^{-4}$ emu mol$^{-1}$ Oe$^{-1}$, $T_{\rm E}$ = 640(10) K, $C$ = 0.01393(7) emu K mol$^{-1}$ Oe$^{-1}$ and $T_{0}$ = 1.38(2) K.
Using the fitted $C$, the calculated effective moment $\mu_{\rm eff}$ of YRu$_{3}$Si$_{2}$ is 0.3338(8) $\mu_{\rm B}$/f.u., corresponding to 0.1113(3) $\mu_{\rm B}$/Ru. This value is very close that in LaRu$_{3}$Si$_{2}$ (0.105 $\mu_{\rm B}$/Ru) \cite{BXLi} and thus such Curie-Weiss behavior could be related to the existence of local moments of Ru atoms in YRu$_{3}$Si$_{2}$.
On the other hand, the Kadawaki-Woods ratio $A/\gamma^{2}$ and the Wilson ratio $R_{\rm W} = {4 \pi^{2} k_{B}^{2} \chi(0)}/{3(g \mu_{B})^{2} \gamma}$ are often used to characterize the strength of electron correlations, where $g$ is Lande factor which takes about 2 for an electron and $\mu_{B}$ is the Bohr magneton \cite{Kadowaki,Wilson}.
Using above values of $A$ = 2.38(7)$\times 10^{-3}$ $\mu \Omega$ cm, $\gamma$ = 27.5(8) mJ mol$^{-1}$ K$^{-2}$ and $\chi(0)$ = 5.56(2)$\times$10$^{-4}$ emu mol$^{-1}$ Oe$^{-1}$, the calculated $A/\gamma^{2}$ and the $R_{\rm w}$ is 3.15(2) $\mu \Omega$ cm mol$^{2}$ K$^{2}$ J$^{-2}$ and 1.49(4), respectively (summarized in Table I).
The value of $A/\gamma^{2}$ is much larger that those in transition metals (0.4 $\mu \Omega$ cm mol$^{2}$ K$^{2}$ J$^{-2}$) and relatively smaller than the universal value 10 $\mu \Omega$ cm mol$^{2}$ K$^{2}$ J$^{-2}$ in heavy-fermion systems \cite{Kadowaki,Jacko}.
The value of $R_{\rm W}$ is larger than 1 that is expected in noninteracting free electron gas. Both large $A/\gamma^{2}$ and $R_{\rm W}$ suggest strong electron correlations in YRu$_{3}$Si$_{2}$.

\section{Conclusion}

In summary, kagome metal YRu$_{3}$Si$_{2}$ shows a superconducting transition at $T_{c}\sim$ 3.0 K. The zero-temperature $\mu_{0}H_{c1}(0)$ and $\mu_{0}H_{c2}(0)$ is 28.0(3) mT and 0.655(2) T, respectively. The derived $\kappa_{\rm GL}$ is 1.11(4), confirming that YRu$_{3}$Si$_{2}$ is a type-II superconductor.
Moreover, the relatively small ${\Delta C_{\rm ele}}/{\gamma T_{c}}$ and $\lambda_{\rm e-ph}$ indicates that YRu$_{3}$Si$_{2}$ has a weakly or moderately coupled BCS bulk superconductivity.
On the other hand, rather strong electron correlation effect is identified by the large values of $A/\gamma^{2}$ and $R_{\rm W}$.
This could be related to the existence of kagome flat band in YRu$_{3}$Si$_{2}$, which can result in high density of states near the $E_{\rm F}$.

\section{Acknowledgements}

This work was supported by National Key R\&D Program of China (Grant No. 2018YFE0202600), Beijing Natural Science Foundation (Grant No. Z200005), the Fundamental Research Funds for the Central Universities and Research Funds of Renmin University of China (RUC) (Grant No. 18XNLG14, 19XNLG13 and 19XNLG17), and Beijing National Laboratory for Condensed Matter Physics.

\end{document}